\documentclass[journal,twoside]{IEEEtran}

\ifCLASSINFOpdf
\usepackage[pdftex]{graphicx}

\else

\fi

\usepackage[cmex10]{amsmath}
\usepackage{amssymb}   
\usepackage{amsxtra}
\usepackage{amscd}
\usepackage{amsthm}
\usepackage{textcomp}
\usepackage{graphicx}  
\usepackage{color}

\usepackage{balance}

\usepackage{array}  

\usepackage{multirow}  

\usepackage{amsmath}
\usepackage{cite}

\usepackage{url}

\usepackage{color,soul} 
\soulregister\cite7
\soulregister\ref7
\soulregister\pageref7

\begin{document}
%

\title{{\huge Communication by Means of Thermal Noise: \\Towards Networks with Extremely Low Power Consumption }}


%
%
%

\author{Ertugrul~Basar,~\IEEEmembership{Fellow,~IEEE}
\thanks{Manuscript received June 22, 2022, revised September 14, 2022, and November 14, 2022; accepted December 5, 2022. Date of publication December X, 2022; date of current version December X, 2022. The associate editor coordinating the review of this article and approving it for publication was S. Jin.}
\thanks{The author is with the Communications Research and Innovation Laboratory (CoreLab), Department of Electrical and Electronics Engineering, Ko\c{c} University, Sariyer 34450, Istanbul, Turkey. (e-mail: ebasar@ku.edu.tr).}
\thanks{MATLAB codes available at https://corelab.ku.edu.tr/tools. }
\thanks{Color versions of one or more figures in this article are available at https://doi.org/10.1109/TCOMM.2022.3228290}
\thanks{Digital Object Identifier 10.1109/TCOMM.2022.3228290}
}

\maketitle

\begin{abstract}

In this paper, the paradigm of \textit{thermal noise communication (TherCom)} is put forward for future wired/wireless networks with extremely low power consumption. Taking backscatter communication (BackCom) and reconfigurable intelligent surface (RIS)-based radio frequency chain-free transmitters one step further, a thermal noise-driven transmitter might enable zero-signal-power transmission by simply indexing resistors or other noise sources according to information bits. This preliminary paper aims to shed light on the theoretical foundations, transceiver designs, and error performance derivations as well as optimizations of two emerging TherCom solutions: Kirchhoff-law-Johnson-noise (KLJN) secure bit exchange and wireless thermal noise modulation (TherMod) schemes. Our theoretical and computer simulation findings reveal that noise variance detection, supported by sample variance estimation with carefully optimized decision thresholds, is a reliable way of extracting the embedded information from noise modulated signals, even with limited number of noise samples. 
  
\end{abstract}
\begin{IEEEkeywords}
Thermal noise communication (TherCom),  Kirchhoff-law-Johnson-noise (KLJN), thermal noise modulation (TherMod), bit error probability.  
\end{IEEEkeywords}

%
\IEEEpeerreviewmaketitle

\section{Introduction}

\begin{figure*}[!t]
	\begin{center}
		\includegraphics[width=1.6\columnwidth]{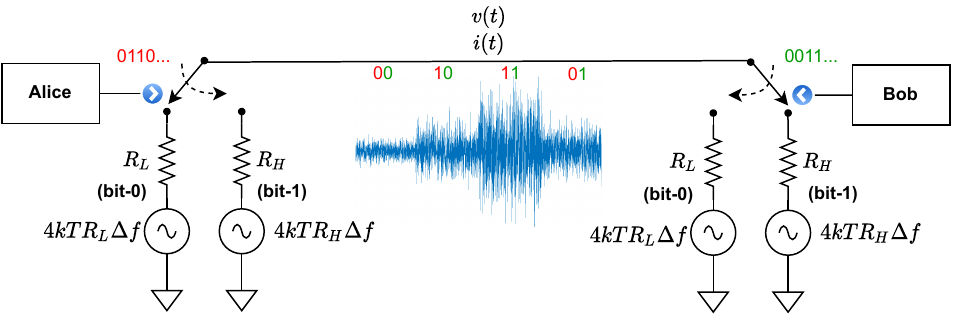}
		\vspace*{-0.3cm}\caption{Wired communication by means of thermal noise: KLJN secure bit (or key) exchange scheme. Three representative thermal noise voltage variances are shown for $00$, $01/10$, and $11$ cases ($00\rightarrow$ low, $01/10\rightarrow$ intermediate, and $11\rightarrow$ high). }\vspace*{-0.3cm}
		\label{fig:KLJN}
	\end{center}
\end{figure*} 

\IEEEPARstart{W}{hile} the first 5G-Advanced standard (3GPP Release 18) is being developed gradually with certain advancements, researchers have already begun exploring radical communication paradigms towards 6G wireless networks of 2030 and beyond. Within this context, reconfigurable intelligent surface (RIS)-empowered communication and backscatter communication (BackCom) have received tremendous interest in the past couple of years \cite{Basar_Access_2019,DiRenzo_2020,Liu_2013,Liu_2019}. While the primary focus of an RIS-assisted system is the manipulation of the end-to-end channel to increase the communication quality-of-service, using an RIS as a transmitter illuminated by an unmodulated radio frequency (RF) source, it might be possible to design transmitters with relatively lower complexity \cite{Basar_2021,Tang_2019}. On the other hand, in BackCom, which is an emerging passive communication paradigm, the primary objective is to convey information by modulating an RF signal generated by another source. Since a BackCom transmitter (tag) does not generate its own RF signals, its power consumption can be as low as a couple of microwatts \cite{Liu_2013}. However, even ambient BackCom systems rely on existing signals, such as TV or Wi-Fi signals, to transmit information. As an alternative to BackCom, communication by means of modulated Johnson (thermal) noise has been recently put forward in \cite{Kapetanovic_2021}, to transmit information with extremely low power consumption and without requiring pre-existing RF signals. In this context, noise-based communication might further reduce the transmitter complexity and power consumption of RIS-based and BackCom systems.

Despite still not being fully perceived by our community, the roots of the concept of communication by means of modulated thermal noise date back to early 2000s. In this paper, we refer to this general paradigm as \textit{thermal noise communication (TherCom)}. In the seminal work of Kish \cite{Kish_2005}, the concept of zero-signal-power (stealth) communication has been put forward by representing an information bit by the choice of two different resistance levels (impedance bandwidths) and the resulting two thermal noise spectra. In the same study, a wireless transmission system with two parabolic antennas is also envisioned to realize the concept of reflection modulation (by indexing a resistor to embed information). Taking stealth communication one step further, the Kirchhoff-law-Johnson-noise (KLJN) secure key exchange scheme is proposed by the same author in 2006 to achieve unconditionally secure communication by utilizing the pure laws of physics: Kirchhoff's law and thermal noises of two pairs of resistors \cite{Kish_2006, Kish_2006_2}. In simple terms, communicating two partners, Alice and Bob, first select their resistors according to information bits and then connect them to a wire channel in each transmission interval, where bit $0$ and bit $1$ are represented by the selection of low- and high-valued resistors, respectively. A secure bit exchange takes place when the bit values at the two ends are different, which results in an intermediate mean-square noise voltage level on the line. Despite the fact that this intermediate level can be detected by an eavesdropper (Eve), the specific contributions of Alice and Bob cannot be comprehended (for Alice$\rightarrow\!\!0$ and Bob$\rightarrow\!\!1$ or vice versa), which ensures an unconditional security that is in the level of quantum secrecy. From a wider perspective, the KLJN scheme has conceptual similarities with the well-known index modulation (IM) concept \cite{Basar_2017,Basar_2019}, which performs indexing to embed information in certain system entities. Specifically, a KLJN communicator performs a sort of IM for the available two resistors, in return, for two noise voltage power spectral densities. 

In \cite{Mingesz_2008}, performance of  KLJN communicators is evaluated through practical experiments over a model-line at several distances and data rates. Furthermore, a bit error rate (BER) of $0.02\%$ is reported through experiments. The effect of wire resistance on the noise voltage and current is investigated in \cite{Kish_2010} and the information leak is reported to be not significant in this case. Additionally, it has been reported in \cite{Gingl_2014} that Johnson-like noise (either generated naturally or externally) must be used for secure key exchange in KLJN systems, while a generalized KLJN scheme is proposed in \cite{Vadai_2015} by using arbitrary (four different) resistors. Nevertheless, in the recent study of \cite{Ferdous_2020}, this generalized scheme is shown to be less secure than the original KLJN scheme under realistic conditions. The first attempt to quantify the bit error probability (BEP) of the KLJN scheme has been made in \cite{Saez_2013}, where an exponentially decaying error probability is reported with respect to the bit duration. In the follow-up study of \cite{Saez_2013b}, the combination of voltage and current measurements is introduced for further BEP reduction. However, in both studies, only the performance of securely exchanged bits is evaluated and approximations, based on the Rice's formula of threshold crossing frequency, are used. A more systematic approach is used in \cite{Smulko_2014} to reveal the effect of distances between noise variances on the BEP performance using statistical hypothesis testing. However, to the best of our knowledge, a general theoretical framework on the receiver system design and BEP optimization of the KLJN scheme have not been presented in the literature.

Against this background, the major contributions of this article is summarized as follows:
\begin{itemize}
	\item By revisiting the scheme of KLJN from a communication engineering perspective, we introduce a new framework for the calculation of its BEP for general system parameters. We also introduce an effective threshold-based detection method that considers sample variance estimation from the taken noise samples.
	\item We propose two novel KLJN detectors by exploiting joint voltage and current measurements to further reduce the BEP. While the first detector raises an error flag when voltage and current bit interpretations are different, the second one adaptively considers either voltage or current measurements according to user information bits.
	\item By taking the KLJN scheme one step further, we propose the scheme of wireless \textit{thermal noise modulation (TherMod)}, which performs a sort of IM for the available two resistors at the transmitter to convey information. We formulate its generic signal model and evaluate the BEP of its proposed detector.
	\item Finally, extensive numerical and computer simulation results are presented to assess the potential of KLJN and TherMod systems under diverse set of parameters. We reveal that the TherCom paradigm might be a remedy for future wired/wireless networks with extremely low or almost zero power consumption.
\end{itemize}

This article is organized as follows. Sections II and III of this manuscript are devoted to theoretical foundations, receiver designs, and BEP optimizations of KLJN and TherMod schemes, respectively. We provide our numerical results under Section IV while concluding the paper in Section V.

\section{Wired Information Transfer with KLJN Secure Bit Exchange} 

In this section, first, we describe the fundamental aspects of the KLJN secure bit exchange scheme from a communication engineering perspective, then introduce a new framework for the calculation of its BEP under general system parameters. Furthermore, we propose two novel detectors to further improve the BEP performance of the KLJN scheme.

\vspace*{-0.3cm}
\subsection{Fundamentals of the KLJN Secure Bit Exchange Scheme}

We begin our discussion by reviewing the KLJN secure bit exchange scheme given in Fig. \ref{fig:KLJN}, which finds its roots in early works of Kish from 2005-2006 \cite{Kish_2005, Kish_2006, Kish_2006_2}. This scheme is based on the Johnson-Nyquist noise (thermal noise, Johnson noise, or Nyquist noise) voltages generated by two terminals, Alice and Bob, which are connected with a wire channel. Here, for each bit duration (bit exchange period) of $T_b$ seconds, according to their information bits, Alice and Bob select one of their resistors with either $R_L$ or $R_H$ ohms. The resistors selected by Alice and Bob are represented by $R_A$ and $R_B$, respectively, that is, $R_A,R_B \in \left\lbrace R_L,R_H  \right\rbrace $. Namely, bit $ 0 $ and bit $ 1 $ are presented by the resistors with low resistance $(R_L)$ and high resistance $(R_H)$, respectively. Here, we can consider the relationship $R_H=\alpha R_L$, where typical $\alpha$ values can range from $5$ to $50$. From this perspective, the KLJN scheme can be considered as an instance of IM, where according to incoming information bits, the index of a resistor is selected at both sides of the link simultaneously. Alternatively, the bit loading process of the KLJN scheme can be considered as the modulation of the noise fluctuations over the channel, where one performs a sort of IM for the noise power spectral density level, that is, for the mean-square noise voltage. Another important aspect of the KLJN communicator is that as in IM schemes, indexing does not consume power and using background noises only, a zero-signal-power communication might be possible without ambient signal sources. The above resistance selection process is repeated in every $T_b$ seconds, where Alice and Bob perform either voltage or current measurements, or both, simply taking samples from randomly fluctuating voltage/current waveforms over the wire at discrete-time instances for further processing.

The mean-square value of the thermal noise voltage $v_R(t)$, which is in fact a Gaussian random process appearing across the terminals of the selected resistor, measured in a bandwidth of $\Delta f $ Hz, is given by \cite{Haykin_2001}
\begin{equation}
	\mathrm{E}[v_R^2(t)]=4kTR\Delta f \, \,\text{volts}^2,
\end{equation}
where $k$ is the Boltzmann's constant, which is $1.38\times 10^{-23}$ joules per degree Kelvin, $T$ is the temperature in degrees Kelvin, and $R\in \left\lbrace R_L,R_H \right\rbrace $ is the selected resistance in ohms. Accordingly, both Alice and Bob rely on voltage $(v(t))$ and/or current $(i(t))$ measurements on the wire line to decode the bits of their partner. Here, considering the Kirchhoff’s law, the power density spectrum of the resulting noise voltage on the line
with two parallel resistors will be
\begin{equation}
	S_{v}(f)=4kT\frac{R_A R_B}{R_A+R_B} \,\, \text{watts/Hz}.
\end{equation}
Similarly, using Ohm's law, the power density spectrum of the noise current flowing through the loop is given by
\begin{equation}\label{eq:3}
	S_{i}(f)=\frac{4kT}{R_A+R_B} \,\, \text{watts/Hz}.
\end{equation}
For the sake of simplicity, we first put our emphasis on voltage measurements, while a generalization will be provided in Subsection II.C. We also note that external noise generators with a relatively high effective temperature (beyond a trillion of Kelvin) and limited bandwidth, in return, much stronger noise signals with the same linear scaling between the resistance and noise voltage spectrum, can be used to further boost the system against background noise effects and attenuation. However, the original thermal noise generation concept can be utilized to realize networks with extremely low power consumption as well as to achieve stealth communication. In other words, the use of background noise signals might be a powerful tool to hide information signals. However, in both cases, the bandwidth of the wire limits the data rate for reliable communication. Finally, it is worth noting that due to thermal equilibrium temperature $T$ at both ends of the link, the mean power flow between the parallel resistors is zero, and voltage and current samples taken any time instant are statistically independent \cite{Kish_2006}.

Against this background, for the following four cases of selected Alice/Bob bits, namely $00,01,10,$ and $11$, where the first and second bits stand for the selected bits by Alice and Bob, respectively, the samples taken from the voltage waveform on the line will be Gaussian distributed with the following variance values:
\begin{align}
	\sigma_{00}^2&= 4kT\frac{R_L R_L}{R_L+R_L} \Delta f=4kT\frac{R_L }{2} \Delta f \nonumber \\
	\sigma_{01}^2&=\sigma_{10}^2= 4kT\frac{R_L R_H}{R_L+R_H} \Delta f=4kTR_L\frac{ \alpha}{1+\alpha} \Delta f \nonumber \\
		\sigma_{11}^2&= 4kT\frac{R_H R_H}{R_H+R_H} \Delta f=4kTR_L\frac{ \alpha}{2} \Delta f.
		\label{eq:variances}
\end{align}  
An interesting interpretation of \eqref{eq:variances} is given as follows \cite{Cho_2005}. As illustrated in Fig. \ref{fig:KLJN}, if both Alice and Both select the large resistance $R_H$, the fluctuations on the line will be high. If both select the small resistance $R_L$, they will be small. And if one user selects the large one while the other selecting the small, the noise variance takes an intermediate value. According to \eqref{eq:variances}, the relative variance ratios are obtained as $1:2\alpha/(1+\alpha):\alpha$, that is, it can be easily shown that for $\sigma_{00}^2=\sigma^2$, we obtain  $\sigma_{01}^2=\sigma_{10}^2=(2\alpha/(1+\alpha))\sigma^2$ and $\sigma_{11}^2=\alpha \sigma^2$. For instance, for $\alpha=10$, which is a value considered in our numerical results, we obtain the variance ratios of $1\!:\!1.8182\!:\!10$, which results in a non-uniform distribution for the observed noise variances. Particularly, as we will discuss later, these noise variance ratios play an important role to distinguish the selected Alice/Bob bit combinations from limited number of measurements (noise samples) and directly affect the overall BEP.

In order to provide unconditional security, KLJN scheme relies on basic laws of physics. Specifically, the KLJN scheme provides a bold answer to the following question: Can Eve find out the selected resistance values at both ends of the link from her own voltage/current measurements? The answer of this question is even more interesting. The eavesdropper can estimate the noise variance on the link from its limited number of samples, while the cases of $01$ and $10$ impose a serious challenge. Specifically, for the cases of $00$ and $11$, which correspond to highest and lowest noise fluctuations on the link, an eavesdropper can easily identify Alice/Bob bits from its measurements, and these bits will be identical undoubtedly. This case is regarded as non-secure bit exchange. Nevertheless, since the cases of $01$ and $10$ produce equivalent noise variances under ideal conditions, that is, with zero wire resistance, the specific locations of $0$ and $1$ bits cannot be determined from measurements taken in-between terminals. This provides a sort of very powerful and simple encryption, and ensures an absolute security \cite{Kish_2016}. Furthermore, the KLJN scheme has been shown be highly resistant to many attacks, including man-in-the-middle, current injection, wire resistance, and cable capacitance attacks. In this sense, from a security performance perspective, one would expect a very high secrecy capacity from the KLJN scheme. On the other hand, Alice and Bob can determine the other partner's selected bit for all four cases exploiting their voltage and/or current samples, thanks to the knowledge of their own bit. In other words, a user's own bit behaves as a decryption mechanism, revealing information about the contribution of the selected resistor at the other end of the link. From this perspective, the KLJN scheme also performs a sort of telecloning of a user's bits, say Alice, allowing Bob to perform detection without physically accessing Alice's bits or transmitted waveforms in his own decision device, and vice versa.

\begin{table*}[!t]
	\centering
	\renewcommand\arraystretch{1.1}
	\caption{Matrix of correct and erroneous decisions for Alice and Bob in KLJN Communication}
	\begin{tabular}{clc|cccc|}
		\cline{4-7}
		\multicolumn{1}{l}{}                             &                                          & \multicolumn{1}{l|}{} & \multicolumn{4}{c|}{Selected Bits (Alice/Bob)}                                                                                                                                                                                                                                                                                                                           \\ \cline{4-7} 
		\multicolumn{1}{l}{}                             &                                          & \multicolumn{1}{l|}{} & \multicolumn{1}{c|}{00}                                                                       & \multicolumn{1}{c|}{01}                                                                       & \multicolumn{1}{c|}{10}                                                                       & 11                                                                       \\ \hline
		\multicolumn{1}{|c|}{\multirow{8}{*}{\vspace*{-2.4cm}Decisions}} & \multicolumn{1}{l|}{\multirow{2}{*}{\vspace*{-0.4cm}00}} & Alice                 & \multicolumn{1}{c|}{Correct}                                                                  & \multicolumn{1}{c|}{\begin{tabular}[c]{@{}c@{}}Error\\ $P_A(01\rightarrow 00)$\end{tabular}}  & \multicolumn{1}{c|}{-}                                                                        & -                                                                        \\ \cline{3-7} 
		\multicolumn{1}{|c|}{}                           & \multicolumn{1}{l|}{}                    & Bob                   & \multicolumn{1}{c|}{Correct}                                                                  & \multicolumn{1}{c|}{-}                                                                        & \multicolumn{1}{c|}{\begin{tabular}[c]{@{}c@{}}Error \\ $P_B(10\rightarrow 00)$\end{tabular}} & -                                                                        \\ \cline{2-7} 
		\multicolumn{1}{|c|}{}                           & \multicolumn{1}{l|}{\multirow{2}{*}{\vspace*{-0.4cm}01}} & Alice                 & \multicolumn{1}{c|}{\begin{tabular}[c]{@{}c@{}}Error\\ $P_A(00\rightarrow 01)$\end{tabular}}  & \multicolumn{1}{c|}{Correct}                                                                  & \multicolumn{1}{c|}{-}                                                                        & -                                                                        \\ \cline{3-7} 
		\multicolumn{1}{|c|}{}                           & \multicolumn{1}{l|}{}                    & Bob                   & \multicolumn{1}{c|}{-}                                                                        & \multicolumn{1}{c|}{Correct}                                                                  & \multicolumn{1}{c|}{-}                                                                        & \begin{tabular}[c]{@{}c@{}}Error \\ $P_B(11\rightarrow 01)$\end{tabular} \\ \cline{2-7} 
		\multicolumn{1}{|c|}{}                           & \multicolumn{1}{l|}{\multirow{2}{*}{\vspace*{-0.4cm}10}} & Alice                 & \multicolumn{1}{c|}{-}                                                                        & \multicolumn{1}{c|}{-}                                                                        & \multicolumn{1}{c|}{Correct}                                                                  & \begin{tabular}[c]{@{}c@{}}Error\\ $P_A(11\rightarrow 10)$\end{tabular}  \\ \cline{3-7} 
		\multicolumn{1}{|c|}{}                           & \multicolumn{1}{l|}{}                    & Bob                   & \multicolumn{1}{c|}{\begin{tabular}[c]{@{}c@{}}Error \\ $P_B(00\rightarrow 10)$\end{tabular}} & \multicolumn{1}{c|}{-}                                                                        & \multicolumn{1}{c|}{Correct}                                                                  & -                                                                        \\ \cline{2-7} 
		\multicolumn{1}{|c|}{}                           & \multicolumn{1}{l|}{\multirow{2}{*}{\vspace*{-0.4cm}11}} & Alice                 & \multicolumn{1}{c|}{-}                                                                        & \multicolumn{1}{c|}{-}                                                                        & \multicolumn{1}{c|}{\begin{tabular}[c]{@{}c@{}}Error\\ $P_A(10\rightarrow 11)$\end{tabular}}  & Correct                                                                  \\ \cline{3-7} 
		\multicolumn{1}{|c|}{}                           & \multicolumn{1}{l|}{}                    & Bob                   & \multicolumn{1}{c|}{-}                                                                        & \multicolumn{1}{c|}{\begin{tabular}[c]{@{}c@{}}Error \\ $P_B(01\rightarrow 11)$\end{tabular}} & \multicolumn{1}{c|}{-}                                                                        & Correct                                                                  \\ \hline
	\end{tabular}
\end{table*}

We conclude this subsection by summarizing the major features of the KLJN secure bit exchange scheme: \\
\textit{ i)} It allows the simultaneous exchange of Alice's and Bob's bits in a single bit period, \\
\textit{ii)} It is unconditionally secure when the selected bits of Alice and Bob are different, that is, for the cases of $01$ and $10$, which occur $50\%$ of the time for uniform bit probabilities. As a result, it can be used to exchange the secure keys of two partners under certain protocols,\\
\textit{iii)} The use of background thermal noise allows a stealth and extremely low power consuming communication, which might be critical for future wired/wireless systems.  

A more detailed discussion and a complete historical perspective on the KLJN scheme can be found in \cite{Kish_2016}.

\subsection{A New Theoretical Framework on the BEP of KLJN}

The KLJN scheme described in the previous subsection, has random bit errors, due to the limited number of samples taken during the specified bit duration. Particularly, Alice and Bob are subject to certain bit errors depending on the selected bit of their partner. As a result, BEP calculation as well as its optimization are not straightforward tasks. Here, we provide a unified view on the BEP behavior of the KLJN scheme.

In Table I, we provide all possible error events for Alice and Bob considering four possible resistance selection scenarios: low-low, low-high, high-low, and high-high, which stand for bit sequences $00,01,10,$ and $11$, respectively. Here, probabilities of corresponding error events for Alice and Bob respectively denoted by $P_A(.)$ and $P_B(.)$. We consider probabilities for all error events, not only involving secure bit exchange, due to the following two reasons. First, under stealth and ultra-low power communication, we are also interested in non-secure bit exchange, that is, the cases of $00$ and $11$ might be exploited as well. From an encryption perspective, under general conditions, it would be still difficult to decrypt messages with $50\%$ compromised bits for long enough keys, such as $256$-bit keys that are widely used in standards and protocols. Second, since a non-secure combination ($00$ and $11$) can be mistaken as a secure combination ($01$ and $10$), and vice versa, focusing only on $00/11\rightarrow01/10$ error events, as in \cite{Saez_2013}, might be misleading in terms of overall BEP.  Nevertheless, by ignoring $01/10\rightarrow00/11$ error events, one can easily obtain a valid BEP expression for the secure bit exchange protocol using our framework as well.

In light of Table I and also considering the bit error symmetries for Alice and Bob\footnote{As will be shown next, we have $P_A(00 \rightarrow 01)=P_B(00\rightarrow 10)$, $P_A(11 \rightarrow 10)=P_B(11\rightarrow 01)$, $P_A(01\rightarrow 00)=P_B(10\rightarrow00)$, and $P_A(10\rightarrow 11)=P_B(01\rightarrow11)$. As a result, Alice and Bob will have the same overall BEP. For notational simplicity, condition terms are not shown in probability expressions of \eqref{eq:Pb}.}, BEP for the KLJN scheme can be simply expressed as
\begin{align}\label{eq:Pb}
	P_b&=\frac{1}{4} \left[ P_A(00 \rightarrow 01 )+ P_A(11 \rightarrow 10 )\right. \nonumber \\ &\hspace{1.5cm}\left. + P_A(01 \rightarrow 00 )+ P_A(10 \rightarrow 11 )  \right],
\end{align}
where it is assumed that $0$s and $1$s are generated uniformly, that is, each selected bit combination in Table I has a probability of $1/4$. However, the BEP values for bit $0$ and bit $1$ might differ in the general case. In what follows, we will investigate these four error event cases using noise variance estimation.

We assume that during each bit duration, both Alice and Bob take samples from the thermal noise voltage on the wire to determine its variance at both sides of the link. Denoting the $k$th independent noise sample by $x_k$, which follows Gaussian distribution with zero mean and $\sigma_i^2$ variance where $\sigma_i^2 \in \left\lbrace \sigma_{00}^2,\sigma_{01}^2,\sigma_{11}^2 \right\rbrace $, that is, $x_k\sim\mathcal{N}(0,\sigma_i^2)$, the noise variance can be estimated as 
\begin{equation}\label{eq:variance}
	\hat{\sigma}^2=\frac{1}{N}\sum_{k=1}^{N} x_k^2.
\end{equation}
Here, $N$ stands for number of noise samples per bit. Assuming the knowledge of the zero mean of the thermal noise, it can be easily shown that the variance estimator of \eqref{eq:variance} is unbiased, that is, $ \mathrm{E}[\hat{\sigma}^2]=\sigma_i^2 $, where $\mathrm{E}[\cdot]$ stands for the expectation. Ideally, $\hat{\sigma}^2$ would follow chi-square distribution, however, for large enough $N$, due to central limit theorem (CLT), we obtain $\hat{\sigma}^2 \sim \mathcal{N}(\sigma_i^2, 2 \sigma_i^4 /N)$. We note that even for $N=50$, $\hat{\sigma}^2$ approximately fits Gaussian distribution and increasing number of noise samples might improve the quality of the variance estimate significantly. 

The band-limited nature of the noise, which can be caused by either the use of external noise generators and/or the band-limited wire channels, puts a hard limit on the number of samples $N$ that can be taken from the channel by Alice and Bob. Assuming a noise bandwidth of $\Delta f$ Hz, the Wiener–Khinchin theorem states that a maximum of $N=2T_b\Delta f $ samples can be taken per bit to ensure statistically independent samples. As a result, for a fixed bandwidth, increasing $N$ increases $T_b$, in return, reduces the bit rate $R_b=1/T_b$. We finally note that the above limit on $N$ also applies to Eve, no matter how strong her signal processing capabilities, ensuring a solid secrecy.

\begin{figure}[!t]
\centering
		\includegraphics[width=0.7\columnwidth]{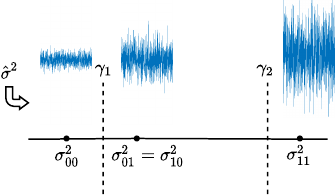}
		\vspace*{-0.3cm}\caption{Threshold-based noise voltage variance detection for the KLJN scheme.}\vspace*{-0.3cm}
		\label{fig:thresholds}
\end{figure} 

As shown in Fig. \ref{fig:thresholds}, we implement a threshold-based detection for the noise variance, where two threshold values $\gamma_1$ and $\gamma_2$ are considered. Here, depending on the value of $\hat{\sigma}^2$, Alice and Bob make decisions on their partner's bit, also exploiting their own bit. Considering statistical decision errors due to the randomness of $\hat{\sigma}^2$, we obtain the corresponding error event probabilities for the cases of $00,11,01,$ and $11$:
\begin{align}\label{eq:Qfunc}
&P_A(00\!\rightarrow\! 01)\!=\! P_B(00\!\rightarrow\! 10)\!=\!P(\hat{\sigma}^2\!>\!\gamma_1) \!=\!Q\left(\!\! \frac{\gamma_1-\sigma_{00}^2}{\sqrt{2\sigma_{00}^4/N}}\!\!\right)\!, \nonumber \\ 	
&P_A(11\!\rightarrow\! 10)\!=\! P_B(11\!\rightarrow\! 01)\!=\!P(\hat{\sigma}^2\!<\!\gamma_2) \!=\!Q\left(\!\! \frac{\sigma_{11}^2-\gamma_2}{\sqrt{2\sigma_{11}^4/N}}\!\!\right)\!, \nonumber \\ 
&P_A(01\!\rightarrow\! 00)\!=\! P_B(10\!\rightarrow\! 00)\!=\!P(\hat{\sigma}^2\!<\!\gamma_1) =Q\left(\!\! \frac{\sigma_{01}^2-\gamma_1}{\sqrt{2\sigma_{01}^4/N}}\!\!\right)\!, \nonumber \\ 
&P_A(10\!\rightarrow\! 11)\!=\! P_B(01\!\rightarrow\! 11)\!=\!P(\hat{\sigma}^2\!>\!\gamma_2) =Q\left(\!\! \frac{\gamma_2-\sigma_{01}^2}{\sqrt{2\sigma_{01}^4/N}}\!\!\right)\!. 
\end{align}
Here, $Q(\cdot)$ denotes the tail probability of the standard Gaussian distribution. Substituting the probability values of \eqref{eq:Qfunc} in \eqref{eq:Pb}, we obtain
\begin{align}\label{eq:8}
P_b&=\frac{1}{4}\left[Q\left( \frac{\gamma_1-\sigma_{00}^2}{\sqrt{2\sigma_{00}^4/N}}\right) +   Q\left( \frac{\sigma_{11}^2-\gamma_2}{\sqrt{2\sigma_{11}^4/N}}\right) \right. \nonumber \\
 &\hspace{0.6cm} \left. + Q\left( \frac{\sigma_{01}^2-\gamma_1}{\sqrt{2\sigma_{01}^4/N}}\right) + Q\left( \frac{\gamma_2-\sigma_{01}^2}{\sqrt{2\sigma_{01}^4/N}}\right)     \right]. 
\end{align}
Let us further simplify \eqref{eq:8} by considering the variance ratios introduced in \eqref{eq:variances}. Noting that $\sigma_{00}^2<\gamma_1 < \sigma_{01}^2 < \gamma_2 < \sigma_{11}^2$, we can simply normalize all these terms with respect to the smallest one, which is $\sigma_{00}^2$, by assuming $\sigma_{00}^2=\sigma^2$. Further defining $\gamma_1=\beta \sigma^2 $ and $\gamma_2=\kappa \sigma^2 $, we obtain 
\begin{align}\label{eq:9}
P_b&=\frac{1}{4}\left[Q\left( \frac{\beta-1}{\sqrt{2/N}}\right) +   Q\left( \frac{\alpha-\kappa}{\alpha\sqrt{2/N}}\right) \right. \nonumber \\
&\hspace{0.6cm} \left. + Q\left( \frac{ \big(\frac{2\alpha}{1+\alpha}\big)-\beta}{\big( \frac{2\alpha}{1+\alpha}\big) \sqrt{2/N}}\right) +  Q\left( \frac{\kappa- \big(\frac{2\alpha}{1+\alpha}\big)}{\big(\frac{2\alpha}{1+\alpha}\big)\sqrt{2/N}}\right)    \right]
\end{align}
where $1 < \beta < \frac{2\alpha}{1+\alpha} <\kappa < \alpha$. It is worth noting that in this ideal transmission scenario, the BEP of \eqref{eq:9} does not depend on the individual noise variances but on the ratio $\alpha$ of two resistance values as well as two thresholds.

In light of these calculations, the BEP of the KLJN scheme can be minimized with respect to the threshold values, that is $\beta$ and $\kappa$, for given $\alpha$ and $N$. At this point, we provide the following remarks to assist our numerical evaluations in Section IV.

\textit{Remark 1}: Even for moderate $N$ and $\alpha$ values, such as $N=50$ and $\alpha=10$, we observe that $P_A(00 \rightarrow 01 ) \gg P_A(11 \rightarrow 10)$ and $P_A(01 \rightarrow 00 ) \gg P_A(10 \rightarrow 11)$ due to uneven distribution of noise variances. In other words, error events associated with the case of $00$ dominates the BEP due to the close proximity of $\sigma_{00}^2$ and $\sigma_{01}^2$. In this case, \eqref{eq:9} can be further simplified as
\begin{equation}\label{eq:10}
P_b \approxeq \frac{1}{4}\left[Q\left( \frac{\beta-1}{\sqrt{2/N}}\right) + Q\left( \frac{ \big(\frac{2\alpha}{1+\alpha}\big)-\beta}{\big( \frac{2\alpha}{1+\alpha}\big) \sqrt{2/N}}\right)    \right]
\end{equation}
which is independent of $\kappa$, that is, the second threshold value. In simple terms, no matter how large $\alpha$ and $\sigma^2$, the case of $00$ becomes the bottleneck of the system.

\textit{Remark 2}: For the case of $R_H \gg R_L$, which is highly practical for the effective modulation of the noise power density spectrum, we have $\alpha \gg 1$ and $ \frac{2\alpha}{1+\alpha}\approx 2$, as a result, \eqref{eq:10} can be further simplified to

\begin{equation}\label{eq:11}
P_b\approx \frac{1}{4}\left[Q\left( \frac{\beta-1}{\sqrt{2/N}}\right) + Q\left( \frac{ 2-\beta}{2 \sqrt{2/N}}\right)    \right].
\end{equation}
As seen from \eqref{eq:11}, only the choice of $\beta$ and $N$ dictates $P_b$. Here $N$ is a measure of the overall quality of the noise variance estimates and directly influences $P_b$ while $\beta$ builds a border between $00$ and $01/10$ decision regions and affects the error probabilities associated with the case of $00$.
 
\textit{Remark 3}: An intuitive solution to the minimization problem of \eqref{eq:11} with respect to $\beta$ would be $\beta=4/3$, which ensures a uniform error probability for bit $0$ and bit $1$ by providing the same error probabilities for the two terms in \eqref{eq:11}. In this case, we obtain     
\begin{equation}\label{eq:12}
P_b\approx \frac{1}{2}Q\left( \frac{1}{3\sqrt{2/N}}\right).
\end{equation}
\eqref{eq:12} reveals an exponentially decaying BEP for the KLJN scheme with respect to $N$, i.e., $P_b\approx (1/24)\exp(-N/36)$ for large $N$ using the exponential approximation for the $Q$-function, which might be promising for wired TherCom schemes. We also note that in the asymptotic case for $N$, the effect of $\beta$ diminishes and $P_b$ is dominated by $N$.

\subsection{New KLJN Detectors}
In this subsection, we propose two new detectors for the KLJN scheme using joint voltage and current measurements. Considering the fact that current and voltage amplitudes are independent due to the second law of thermodynamics \cite{Saez_2013b}, the BEP of the KLJN scheme can be further improved with these detectors by jointly processing voltage and current samples. We first provide the basics of current-based noise variance estimation and then present the two new detectors.

\begin{figure}[!t]
	\centering
	\includegraphics[width=0.7\columnwidth]{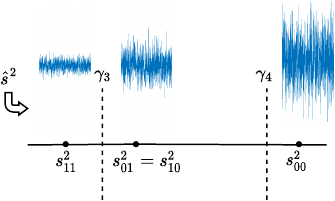}
	\vspace*{-0.3cm}\caption{Threshold-based noise current variance detection for the KLJN scheme.}\vspace*{-0.3cm}
	\label{fig:thresholds2}
\end{figure}

Considering the power density spectrum of the loop noise current from \eqref{eq:3}, we obtain the following noise variances for the cases of $00,01/10,$ and $11$, respectively:
\begin{align}
 s_{00}^2&= 4kT\frac{1}{R_L+R_L} \Delta f=4kT\frac{1}{2R_L } \Delta f=s^2 \nonumber\\
 s_{01}^2&\!=s_{10}^2\!= 4kT\frac{1}{R_L+R_H} \Delta f\!=4kT\frac{1}{(1+\alpha)R_L} \Delta f\!=\!\frac{2s^2}{1+\alpha} \nonumber\\
s_{11}^2&= 4kT\frac{1}{R_H+R_H} \Delta f=4kT\frac{ 1}{2\alpha R_L} \Delta f=\frac{s^2}{\alpha}.
	\label{eq:variances2}
\end{align} 
Here, noise variance ratios are obtained as $1\!:\!2/(1+\alpha)\!:\!1/\alpha$, which is the opposite of the case for voltage variances, that is, we observe the largest fluctuations for the case of $00$. As a result, we consider the threshold-based modified detection scheme in Fig. \ref{fig:thresholds2} for the case of noise current samples, where $\hat{s}^2$ stands for the unbiased noise variance estimate that is defined the same as in \eqref{eq:variance}.

Following a similar methodology as for voltage measurements, the BEP for the case of current measurements is obtained as 
\begin{align}\label{eq:Pb_current}
	\tilde{P}_b&=\frac{1}{4} \left[ \tilde{P}_A(00 \rightarrow 01 )+ \tilde{P}_A(11 \rightarrow 10 )\right. \nonumber \\ &\hspace{1.5cm}\left. + \tilde{P}_A(01 \rightarrow 00 )+ \tilde{P}_A(10 \rightarrow 11 )  \right],
\end{align}
where $\tilde{P}(\cdot)$ is used here to distinguish from the probabilities of voltage-based error events. Defining $\gamma_3=\eta s^2$ and $\gamma_4=\xi s^2$, considering $\hat{s}^2 \sim \mathcal{N}(s_i^2,2s_i^4/N)$ for $s_i^2 \in \left\lbrace s_{00}^2,s_{01}^2,s_{11}^2 \right\rbrace $ and following similar steps from the previous subsection, we obtain
\begin{align}\label{eq:15}
	\tilde{P}_b&=\frac{1}{4}\left[Q\left( \frac{1-\xi}{\sqrt{2/N}}\right) +   Q\left( \frac{\eta-\left( \frac{1}{\alpha}\right) }{\left( \frac{1}{\alpha}\right) \sqrt{2/N}}\right) \right. \nonumber \\
	&\hspace{0.4cm} \left. + Q\left( \frac{ \xi-\big(\frac{2}{1+\alpha}\big)}{\big( \frac{2}{1+\alpha}\big) \sqrt{2/N}}\right) +  Q\left( \frac{ \big(\frac{2}{1+\alpha}\big)-\eta}{\big(\frac{2}{1+\alpha}\big)\sqrt{2/N}}\right)    \right].
\end{align}
For large enough $\alpha$ and considering the dominance of $11$ error events due to the close proximity of $s_{11}^2$ and $s_{01}^2$ for this case, \eqref{eq:15} can be simplified as
\begin{equation}\label{eq:16}
	\tilde{P}_b\approxeq \frac{1}{4}\left[ Q\left( \frac{\eta-\left( \frac{1}{\alpha}\right) }{\left( \frac{1}{\alpha}\right) \sqrt{2/N}}\right) + Q\left( \frac{ \big(\frac{2}{\alpha}\big)-\eta}{\big(\frac{2}{\alpha}\big)\sqrt{2/N}}\right)    \right].
\end{equation}
To ensure uniform error probability for bit $0$ and bit $1$, we can set $\eta=4/(3\alpha)$ in \eqref{eq:16} to obtain $\tilde{P}_b\approx \frac{1}{2}Q\big( \frac{1}{3\sqrt{2/N}}\big)$, which is the same as \eqref{eq:12}. Consequently, due to the symmetry in their variance ratios, we observe that voltage and current measurements provide an identical BEP. Nevertheless, by exploiting their independence, we introduce the following two detectors to further improve the error performance.

\begin{figure*}[!t]
	\begin{center}
		\includegraphics[width=1.75\columnwidth]{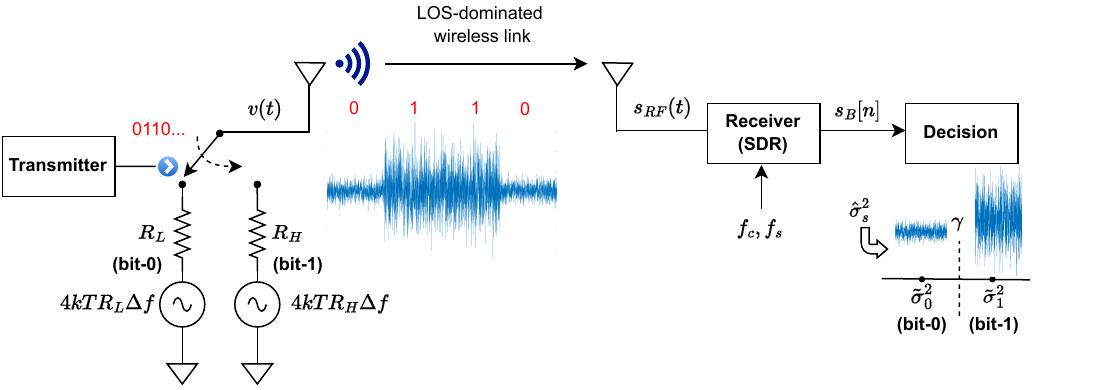}
		\vspace*{-0.3cm}\caption{Wireless communication by means of thermal noise: TherMod scheme. Two representative thermal noise voltage variances are shown for bit $0$ and bit $1$ cases ($0\rightarrow$ low and $1\rightarrow$ high). }\vspace*{-0.3cm}
		\label{fig:TherMod}
	\end{center}
	
\end{figure*} 

\textit{KLJN New Detector I (ND-I):} Taking samples from both voltage and current noise waveforms, this detector raises an error flag when voltage and current bit decisions are different. It is worth noting that a similar detector is considered in \cite{Saez_2013b}, which performs secure bit exchange when both current and voltage bit interpretations are secure ($01/10$). For ND-I, the correct symbol detection probability of Alice can be obtained as follows by considering the same decisions from voltage and current measurements, respectively for $00,11,01,$ and $10$:
\begin{align}\label{eq:17}
	P_c&=\frac{1}{4}  \left[ P(\hat{\sigma}^2<\gamma_1) \, P(\hat{s}^2>\gamma_4)  +  P(\hat{\sigma}^2>\gamma_2) \, P(\hat{s}^2<\gamma_3) \right. \nonumber \\
	&\left.  + P(\hat{\sigma}^2>\gamma_1) \, P(\hat{s}^2<\gamma_4) + P(\hat{\sigma}^2<\gamma_2) \, P(\hat{s}^2>\gamma_3)  \right]. 
\end{align}
\vspace*{-0.35cm}\\ 
Substituting corresponding probabilities in \eqref{eq:17}, we obtain
 \begin{align}\label{eq:18}
 	P_c=&\frac{1}{4}  \left[ Q\bigg( \frac{1-\beta}{\sqrt{2/N}}\bigg)  Q\bigg( \frac{\xi-1}{\sqrt{2/N}}\bigg) \right. \nonumber \\ 
 	 & \,\, +  Q\bigg( \frac{\kappa-\alpha}{\alpha\sqrt{2/N}}\bigg) Q\bigg( \frac{\left( \frac{1}{\alpha}\right) -\eta }{\left( \frac{1}{\alpha}\right) \sqrt{2/N}}\bigg) \nonumber \\ 
 	& \,\, + Q\left( \frac{\beta- \big(\frac{2\alpha}{1+\alpha}\big)}{\big( \frac{2\alpha}{1+\alpha}\big) \sqrt{2/N}}\right)  Q\left( \frac{ \big(\frac{2}{1+\alpha}\big)-\xi}{\big( \frac{2}{1+\alpha}\big) \sqrt{2/N}}\right) \nonumber \\ 
 	& \,\, \left. + Q\left( \frac{ \big(\frac{2\alpha}{1+\alpha}\big)-\kappa}{\big(\frac{2\alpha}{1+\alpha}\big)\sqrt{2/N}}\right) Q\left( \frac{\eta- \big(\frac{2}{1+\alpha}\big)}{\big(\frac{2}{1+\alpha}\big)\sqrt{2/N}}\right)  \right]. 
 \end{align}
Finally, the BEP of this detector is obtained as $P_b^{I}=0.5(1-P_c)$, where the factor of $0.5$ stands for the ratio of bit errors with respect to a symbol error of the KLJN scheme. We observe that the BEP of this detector is a function of $\beta,\xi,\kappa,$ and $\eta$, while a straightforward solution for the optimum set of these parameters that minimizes $P_b^{I}$ cannot be obtained analytically. As a result, for given $\alpha$ and $N$, we minimize $P_b^{I}$ using numerical methods in Section IV.

An important aspect of ND-I is its ability of error detection, which occurs when voltage and current bit interpretations are different. We will show later by numerical results that this detector is extremely robust to bit errors when erroneously detected symbols are discarded.

\textit{KLJN New Detector II (ND-II):} Taking the previous detector one step forward, ND-II considers the fact that $00$ and $11$ error events are the dominant ones for voltage and current measurements, respectively. Consequently, this detector allows Alice and Bob to select their measurement types depending on their own bits. In other words, Alice (or Bob) considers current measurements if its own bit is $ 0 $, while uses voltage measurements if its own bit is $ 1 $, since $00 \leftrightarrow 01$ and $10\leftrightarrow11$ error events are less likely for current and voltage measurements of Alice, respectively. 

In light of this information, Alice and Bob will make their decisions according to the procedures in Table II. It is worth noting that for this detector, Alice and Bob can make different decisions for the cases of $01$ and $10$, but this is very unlikely. Accordingly, the BEP of ND-II is obtained as

\begin{table}[!t]
	\centering
	\renewcommand\arraystretch{1.4}
	\caption{Adaptive Decision Types for KLJN New Detector II}
\begin{tabular}{c|cccc|}
	\cline{2-5}
	& \multicolumn{4}{c|}{Cases}                                                                                                                                                                                                                                                                                                                                                                                                                                                    \\ \cline{2-5} 
	& \multicolumn{1}{c|}{$ 00 $}                                                                                                         & \multicolumn{1}{c|}{$ 11 $}                                                                                        & \multicolumn{1}{c|}{$ 01 $}                                                                                                & $ 10 $                                                                                                \\ \hline
	\multicolumn{1}{|c|}{\!Alice\!}                                                   & \multicolumn{1}{c|}{C}                                                                                                          & \multicolumn{1}{c|}{V}                                                                                         & \multicolumn{1}{c|}{C}                                                                                                 & V                                                                                                 \\ \hline
	\multicolumn{1}{|c|}{\!Bob\!}                                                     & \multicolumn{1}{c|}{C}                                                                                                          & \multicolumn{1}{c|}{V}                                                                                         & \multicolumn{1}{c|}{V}                                                                                                 & C                                                                                                 \\ \hline 
	\multicolumn{1}{|c|}{\begin{tabular}[c]{@{}c@{}}\!Error\! \\ \!Events\!\end{tabular}} & \multicolumn{1}{c|}{\begin{tabular}[c]{@{}c@{}}$\!\tilde{P}_A(00\!\rightarrow\! 01)\!$ \\ $\!\tilde{P}_B(00\!\rightarrow\! 10)\!$\end{tabular}} & \multicolumn{1}{c|}{\begin{tabular}[c]{@{}c@{}}$\!P_A(11\!\rightarrow\! 10)$\!\\ $\!P_B(11\!\rightarrow\! 01)\!$\end{tabular}} & \multicolumn{1}{c|}{\begin{tabular}[c]{@{}c@{}}$\!\tilde{P}_A(01\!\rightarrow\! 00)\!$\\ $\!P_B(01\!\rightarrow\! 11)\!$\end{tabular}} & \begin{tabular}[c]{@{}c@{}}$\!P_A(10\!\rightarrow\! 11)\!$\\ $\!\tilde{P}_B(10\!\rightarrow\! 00)\!$\end{tabular} \\ \hline 
\end{tabular}\vspace*{0.2cm}
\textit{\hspace*{-3cm}C: current measurement, V: voltage measurement}
\end{table}

\begin{align}\label{eq:Pb_II}
	P_b^{II}&=\frac{1}{4} \left[ \tilde{P}_A(00 \rightarrow 01 )+ P_A(11 \rightarrow 10 )\right. \nonumber \\ &\hspace{1.5cm}\left. + \tilde{P}_A(01 \rightarrow 00 )+ P_A(10 \rightarrow 11 )  \right].
\end{align}
Substituting the corresponding probabilities in \eqref{eq:Pb_II}, we obtain
\begin{align}\label{eq:20}
	P_b^{II}&=\frac{1}{4}\left[Q\left( \frac{1-\xi}{\sqrt{2/N}}\right) +   Q\left( \frac{\alpha-\kappa}{\alpha\sqrt{2/N}}\right) \right. \nonumber \\
	&\hspace{0.3cm} \left. + Q\left( \frac{ \xi-\big(\frac{2}{1+\alpha}\big)}{\big( \frac{2}{1+\alpha}\big) \sqrt{2/N}}\right) + Q\left( \frac{\kappa- \big(\frac{2\alpha}{1+\alpha}\big)}{\big(\frac{2\alpha}{1+\alpha}\big)\sqrt{2/N}}\right)    \right].
\end{align}
As seen from \eqref{eq:20}, $P_b^{II}$ does not depend on fragile thresholds, $\beta$ and $\eta$, and only consists of weaker probability terms that are ignored earlier in BEP calculations. However, we again resort numerical tests to determine optimum $\xi$ and $\kappa $ values that minimizes $P_b^{II}$. That is being said, the search space can be reduced by assuming $\xi=\kappa/\alpha$, which provides a uniform error probability for bit $0$ and bit $1$ by ensuring $ \tilde{P}_A(00 \rightarrow 01 )= P_A(11 \rightarrow 10 ) $ and $ \tilde{P}_A(01 \rightarrow 00 )= P_A(10 \rightarrow 11 ) $.

\section{Wireless Information Transfer with Thermal Noise Modulation}
In this section, taking the KLJN scheme one step further, we introduce the scheme of TherMod, which considers the modulation of the thermal noise level to transmit information over wireless channels.

\vspace*{-0.3cm}
\subsection{Fundamentals of Thermal Noise Modulation (TherMod)}

In Fig. \ref{fig:TherMod}, we present the generic block diagram of the TherMod transceiver. Here, similar to the KLJN scheme, incoming information bits at the transmitter determine the index of the selected resistor, in return, the power spectral density of the generated thermal noise waveform\footnote{We note that an alternative transmission scheme can be obtained by simply performing on-off keying for a resistor according to information bits, that is, choosing between a resistor and an open (or short) circuit connection to an antenna, as envisioned in \cite{Kish_2005} and experimentally demonstrated in \cite{Kapetanovic_2021} recently. However, we might expect almost the same error performance for sufficiently high $\alpha$ and to be consistent with our earlier coverage, we consider the generic model of Fig. \ref{fig:TherMod}.}. However, unlike the KLJN scheme, only one of the terminals performs this resistor index selection and an unguided transmission medium is considered. Due to the  extremely low noise power levels at the transmitter, possibly a short-range and line-of-sight (LOS)-dominated communication might be possible, with the aid of high-gain antennas, such as horn antennas, as experimentally demonstrated in \cite{Kapetanovic_2021} for the first time. Here, circuit sensitivity might be one of the most critical practical challenges for this system since the received noise power should be strong enough to be detected. Alternatively, external noise generators can be used to extend the coverage, with the cost of higher transmitter complexity and power consumption. Another way of boosting the noise level over the channel might be employing a power amplifier prior to transmission, which also increases the power consumption. In what follows, we put our emphasis on the basic model of Fig. \ref{fig:TherMod}. We also note that the TherMod scheme does not need and modulate ambient RF signals as in BackCom and differs from classical communication systems that rely on carrier signal modulation.

Denoting the random voltages stemming from the thermal noise of low- and high-valued resistors by $v_L(t)$ and $v_H(t)$, respectively, the transmitted signal can be expressed as 
\begin{equation}
	v(t)=\sum_{n} g(t-nT_b)v_{i_n}(t)
\end{equation}
where $i_n\in \left\lbrace L,H \right\rbrace $ according to the incoming bits. Here, $T_b$ is the bit duration as defined before and $g(t)$ stands for the unit-gain rectangular pulse shape with a duration of $T_b$ seconds. A realization of this waveform, which is also a Gaussian random process with two variance levels, is illustrated in Fig. \ref{fig:TherMod}. Focusing on a single bit duration, the noise samples taken from this process will be independent and identically distributed Gaussian random variables. However, their variance carries information as in the KLJN scheme. Although the thermal noise generated by the resistors is white and can be observed at any frequency, the considered antennas as well as the receiver equipment (sampling rate) limits the bandwidth of the observed noise signal.

Considering a pure LOS link between the transmitter and the receiver without any multipath components and ignoring time delays, according to the Friis transmission equation \cite{Goldsmith}, the received signal is obtained as
 \begin{equation}
 	s_{RF}(t)=\frac{\sqrt{G_tG_r}\lambda}{4\pi d} v(t)
 \end{equation}
where $d$ is the distance between terminals, $G_t/G_r$ are transmit/receive antenna gains, and $\lambda$ is the wavelength. A software-defined radio (SDR)-based receiver with carrier frequency $f_c$ and sampling rate $f_s$ can be used to obtain complex baseband samples for further processing. In light of this information, received complex baseband signal is obtained as $s_B(t)=\text{LPF}\left\lbrace s_{RF}(t)e^{-j2\pi f_c t}\right\rbrace $, where its sampled version can be denoted by $s_B[n]$ (or simply by $s_n$). Here, $\text{LPF}$ stands for low-pass filtering after downconversion. Filtering operations within the SDR ensure a band-limited white noise process at the output with independent in-phase and quadrature components. As a result, the $n$th sample in the complex baseband can be represented as\footnote{In case of fading, the system can still operate using clever algorithms instead of threshold-based detection if the channel coefficient remains the same for $T_b$ seconds. For the architecture of Fig. 4, classical channel estimation might not be possible since we cannot transmit pilot signals.} 
\begin{equation}\label{eq:TNM}
s_n=r_n + w_n
\end{equation}
where $r_n$ stands for the samples of the received useful signal stemming from the thermal noise generated by the transmitter while $w_n$ is the additive while Gaussian noise (AWGN) sample introduced due to the receiver circuitry (amplification, filtering, and downconversion). In other words, received complex baseband samples include not only the information carrying noise terms but also the disruptive noise terms added on top of them. In simple terms, the variance of $w_n$ is fixed and proportional to $\sigma_w^2 \sim kTB$, where $B\approx f_s$ according to the Nyquist theorem. On the other hand, the variance of $r_n$ changes with respect to the selected resistor at the transmitter and proportional to $\sigma_r^2 \sim 4kTR_i B P_G$ for $i\in \left\lbrace L,H \right\rbrace $, where $P_G$ stands for the overall path gain including free space propagation loss and antenna gains, i.e., $P_G \propto (\frac{\lambda}{4 \pi d})^2 G_t G_r$. As shown in Fig. 4, the primary task of the receiver is to make a decision on the transmitted bit by processing the received samples ($s_n$) accordingly.

Considering this basic communication model, in the following subsection, we provide a theoretical framework to assess the BEP performance of the TherMod scheme.

\subsection{A Theoretical Perspective on TherMod}
In this subsection, considering the statistics of the received noise samples as well as the AWGN terms at the receiver, we derive the theoretical BEP of the TherMod scheme.

As in the KLJN scheme, rather than the individual noise variances, their  ratio dictates the overall BEP. Denoting the variance of $r_n$ for bit $0$ $(R_L)$ and bit $1$ $(R_H)$ by $\sigma_0^2$ and $\sigma_1^2$, respectively, we have $\sigma_1^2 = \alpha \sigma_0^2 $, where $\alpha $ is as defined before and stands for the ratio of two resistance values. To assess the BEP, we define a quality metric, similar to the signal-to-noise ratio (SNR) in traditional communication systems, by $\delta= \sigma_0^2/ \sigma_w^2 $. Here, $\delta$ relates the useful (information carrying) noise variance to the receiver (disruptive) noise variance and directly affects the error performance. In light of this information and considering \eqref{eq:TNM}, we obtain $s_n \sim \mathcal{CN}(0,\tilde{\sigma}_0^2)$ and $s_n \sim \mathcal{CN}(0,\tilde{\sigma}_1^2)$  for bit $ 0 $  and bit $ 1 $, respectively, where $\tilde{\sigma}_0^2=\sigma_w^2(1+\delta)$ and $\tilde{\sigma}_1^2=\sigma_w^2(1+\alpha\delta)$. Here $\mathcal{CN}(0,\sigma^2)$ stands for the complex Gaussian distribution with zero mean and $\sigma^2$ variance.

Similar to the KLJN scheme, we consider sample variance calculations using $N$ complex baseband samples for each bit duration, as a result, a total of $f_s=N/T_b$ complex samples are processed per second. Considering the complex nature of $s_n$, its variance can be estimated as
\begin{equation}
\hat{\sigma}_s^2=\frac{1}{N} \sum_{n=1}^{N} \left| s_n \right|^2. 
\end{equation}
Here, for large enough $N$, the distribution of $\hat{\sigma}_s^2$ can be approximated as Gaussian and follows $\mathcal{N}(\tilde{\sigma}_i^2,\tilde{\sigma}_i^4/N)$ for $i\in \left\lbrace 0,1 \right\rbrace $. Accordingly, the BEP of the TherMod scheme is obtained as
\begin{equation}
	P_b=\frac{1}{2} \left[ P(0\rightarrow 1 \left. \right| 0 ) + P(1\rightarrow 0 \left. \right| 1 ) \right]. 
\end{equation}
Considering a threshold-based variance detection as shown in Fig. 4, where a decision is made on the transmitted bit considering a predetermined threshold $\gamma$ for $\hat{\sigma}_s^2$, we obtain
\begin{align}\label{eq:26}
P_b&=\frac{1}{2} \left[ P(\hat{\sigma}_s^2 >\gamma \left. \right| 0 ) + P(\hat{\sigma}_s^2<\gamma \left. \right| 1 ) \right] \nonumber \\
&= \frac{1}{2} \left[ Q\left( \frac{\gamma-\tilde{\sigma}_0^2}{\sqrt{\tilde{\sigma}_0^4/N}} \right) + Q\left( \frac{\tilde{\sigma}_1^2-\gamma}{\sqrt{\tilde{\sigma}_1^4/N}} \right)  \right]. 
\end{align}
Scaling the threshold with respect to $\sigma_w^2$ as $\gamma=\chi\sigma_w^2$ for $1+\delta < \chi < 1+\alpha\delta$, \eqref{eq:26} can be re-expressed as 
\begin{equation}\label{eq:27}
	P_b=\frac{1}{2}\left[ Q\left( \frac{\chi-(1+\delta)}{(1+\delta)\sqrt{1/N}} \right) + Q\left( \frac{(1+\alpha\delta)-\chi}{(1+\alpha\delta)\sqrt{1/N}} \right)  \right].
\end{equation}
Here, we observe that the BEP of the TherMod scheme depends on the selected threshold $(\chi)$, the ratio of resistance values $(\alpha)$ and the ratio of useful and disruptive noise variances $(\delta)$. In what follows, we further investigate \eqref{eq:27} from different perspectives.

\textit{Remark 4}: To have uniform error probabilities for bit $ 0 $ and bit $ 1 $, we can equate the arguments of $Q$-functions in \eqref{eq:27} by
\begin{equation}\label{eq:28}
	\chi= \frac{2(1+\delta)(1+\alpha\delta)}{2+\delta(1+\alpha)}
\end{equation} 
for which \eqref{eq:27} simplifies to
\begin{equation}\label{eq:29}
	P_b=Q\left(\frac{\sqrt{N}\delta(\alpha-1)}{2+\delta(1+\alpha)} \right). 
\end{equation}
We note that $\chi$ given in \eqref{eq:28} is valid for $\alpha >1$, which is also a requirement for the operation of TherMod. For the case of $\alpha \gg 1$, we obtain 
\begin{equation}\label{eq:30}
	P_b\approx Q\left(\frac{\sqrt{N}\alpha\delta}{2+\alpha\delta} \right) 
\end{equation}
which reveals an exponentially decaying BEP with respect to $N$. This can be also verified for the special case of $\alpha\delta >> 1$ for which one obtains $P_b\sim Q (\sqrt{N})$ from \eqref{eq:30}. On the other hand, for small $\delta$, the gap between $\tilde{\sigma}_0^2$ and $ \tilde{\sigma}_1^2 $ will shrink and one can obtain $P_b \sim Q(\sqrt{N} \alpha\delta/2 ) $ for $\alpha\delta \ll 1$, which results in a degraded BEP performance compared to the earlier case.

\section{Numerical Results}

In this section, we provide our computer simulation and numerical results for both KLJN and TherMod schemes. For these two schemes, we assume that the ratio of high- and low-valued resistances is $\alpha=10$ unless specified otherwise.

\subsection{Results for the KLJN Scheme}

\begin{figure}[!t]
	\centering
	\includegraphics[width=1\columnwidth]{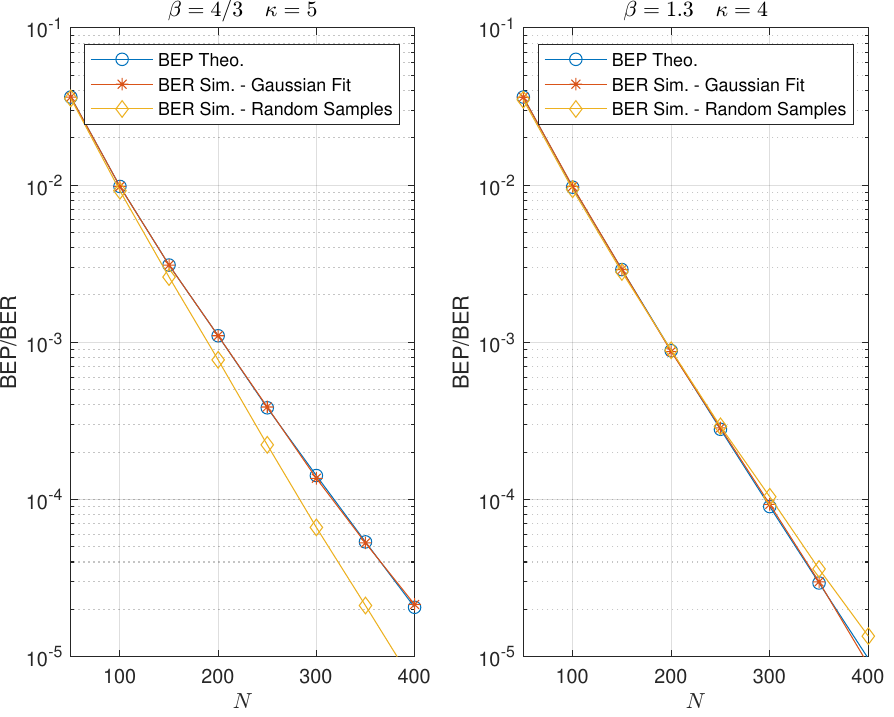}
	\vspace*{-0.6cm}\caption{Comparison of theoretical BEP and computer simulation BER results for two different set of threshold values in the KLJN scheme.}\vspace*{-0.3cm}
	\label{fig:Fig5}
\end{figure}

In Fig. \ref{fig:Fig5}, we plot the BEP performance of the KLJN scheme using voltage-based measurements only, where we assumed $\beta=4/3,\kappa=5$ and $\beta=1.3,\kappa=4$. Here, we observe that when we model the sample variance directly as $\hat{\sigma}^2 \sim \mathcal{N}(\sigma_i^2, 2 \sigma_i^4 /N)$ during Monte Carlo simulations (Gaussian Fit), theoretical and simulation results perfectly match. On the other hand, BEP results obtained by generating $N$ Gaussian samples for sample variance calculation according to \eqref{eq:variance} in simulations, slightly deviate from theoretical ones, due to the insufficient fit of chi-square distribution to Gaussian distribution. We also note that this deviation depends on the selected parameters. Despite the fact that even $N=50$ would be a sufficient Gaussian fit for chi-square distribution, the overall convergence is not very fast due to its high skewness \cite{Box_2005}. Nevertheless, our theoretical derivation from \eqref{eq:9} might still be a good indicator for practical BER performance. 

\begin{figure}[!t]
	\centering
	\includegraphics[width=1\columnwidth]{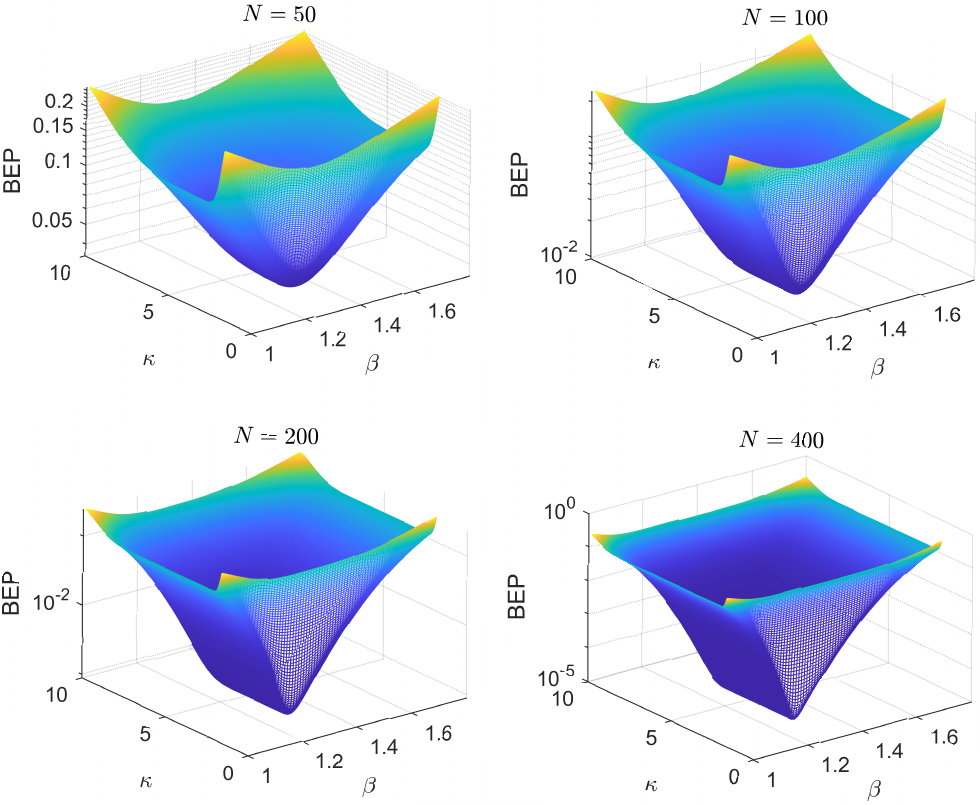}
	\vspace*{-0.6cm}\caption{3D search for minimum BEP over $\kappa$ and $\beta$ (threshold) values for varying number of noise voltage samples.}\vspace*{-0.3cm}
	\label{fig:Fig6}
\end{figure}

In Fig. \ref{fig:Fig6}, we perform a 3D search over $\beta$ and $\kappa$ considering the BEP values obtained from \eqref{eq:9}. Here, we simply vary $N$ from $50$ to $400$. As seen from Fig. \ref{fig:Fig6}, the optimal $\beta$ that minimizes the BEP lies around $1.3$, while the effect of $\kappa$ is not significant at this specific $\beta$ region, which is consistent with our discussion under \textit{Remark 1}. Similar to Fig. \ref{fig:Fig5}, we observe a significant BEP improvement with increasing $N$.

\begin{figure}[!t]
	\centering
	\includegraphics[width=0.85\columnwidth]{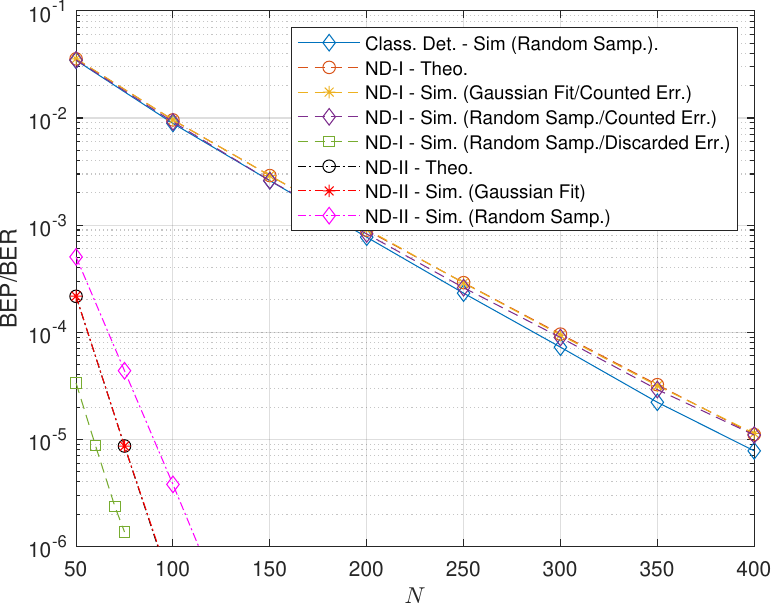}
	\vspace*{-0.3cm}\caption{BEP/BER comparison of classical (voltage-based) KLJN detector and two new KLJN detectors with optimized thresholds for $N=100$.}\vspace*{-0.3cm}
	\label{fig:Fig7}
\end{figure}

In Fig. \ref{fig:Fig7}, we compare the BER performances of our two new detectors with the classical (voltage-based) KLJN detector. For the optimization of the thresholds, we considered $N=100$ for all three detectors with a search resolution between $0.05-0.001$, which provided $\beta=1.3160,\kappa=3.1512$ for the KLJN classical detector, $\beta=1.3150,\kappa=3.1532,\eta=0.1300,\xi=0.3168$ for ND-I and $\kappa=3.1512,\xi=0.3148$ for ND-II. Unsurprisingly, almost the same optimal values are obtained for these detectors thanks to the common probability terms in \eqref{eq:9}, \eqref{eq:18}, and \eqref{eq:20}. We also note that optimal threshold values slightly change with respect to $N$; however, the same thresholds given above are used for all considered $N$ values for simplicity.

As seen from Fig. \ref{fig:Fig7}, ND-I achieves a very similar performance to the classical detector, for the case where random bit errors counted for different voltage and current interpretations, that is, when ND-I raises an error flag. On the other hand, if the corresponding bits are discarded when an error is detected, the performance of ND-I improves splendidly. This behavior can be explained by the fact that it is very unlikely to have bit errors for both voltage and current measurements at the same time. Nevertheless, these discarded bits correspond to the loss of $7$ - $3.5\%$ of the transmitted bits for the considered $N$ values in Fig. \ref{fig:Fig7}, which are between $50$ and $75$. ND-II eliminates the need for error monitoring with a slight degradation in BER compared to ND-I. Thanks to its adaptive sample variance calculation ability among voltage and current samples, a remarkable BER performance is obtained for the ND-II. We note that for all three detectors, our theoretical derivations (respectively given by \eqref{eq:9}, \eqref{eq:18}, and \eqref{eq:20}) are accurate when we fit the sample variance ($\hat{\sigma}^2$ or $\hat{s}^2$) directly to the Gaussian distribution (Gaussian Fit), while a slight gap is observed between theoretical and computer simulation curves when Gaussian noise samples are used for sample variance calculation (Random Samples) due to the insufficient fit of Gaussian distribution to chi-square distribution as discussed earlier. This gap is more visible for ND-II due to its limited number of noise samples and considerably low BEP values. Nevertheless, our theoretical framework based on the Gaussian approximation of the sample variance, stands out as a solid baseline for not only theoretical evaluation but also threshold optimization of the KLJN scheme.

\begin{figure}[!t]
	\centering
	\includegraphics[width=1\columnwidth]{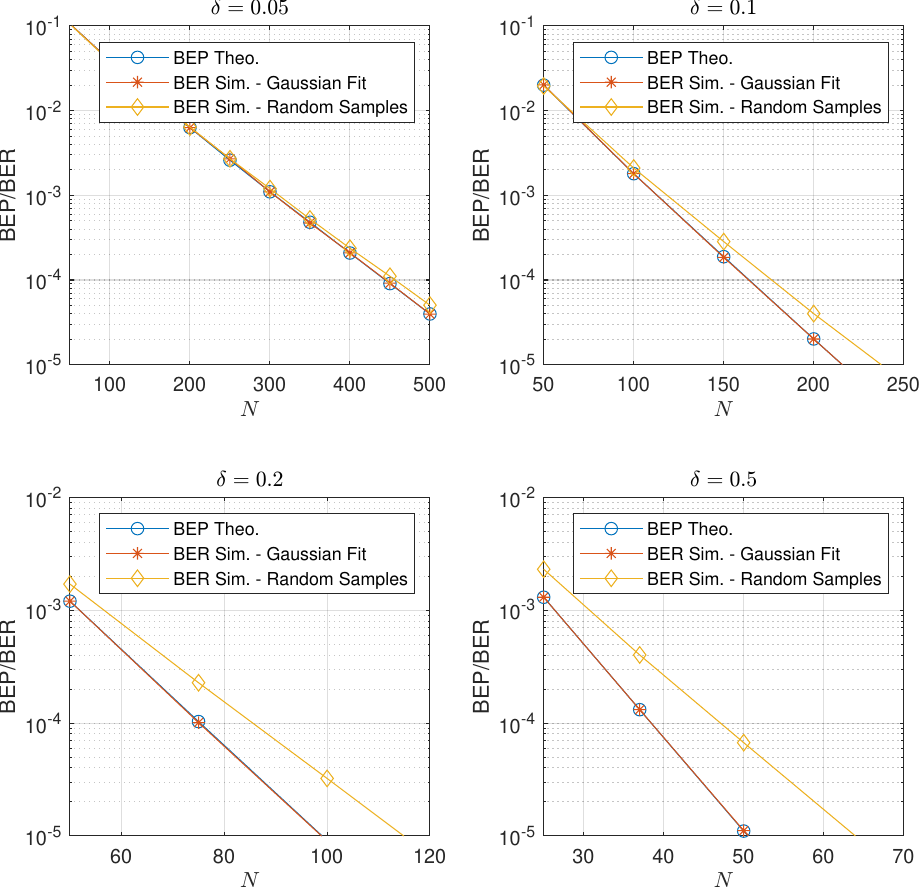}
	\vspace*{-0.6cm}\caption{Comparison of theoretical BEP and computer simulation BER results for different $\delta= \sigma_0^2/ \sigma_w^2 $ values of the TherMod scheme.}\vspace*{-0.3cm}
	\label{fig:Fig8}
\end{figure}

\subsection{Results for the TherMod Scheme}
In Fig. \ref{fig:Fig8}, we provide the BEP performance of the TherMod scheme with respect to number of complex samples per bit ($N$) using the threshold value ($\chi$) obtained in \eqref{eq:28} for varying $\delta$, where $\delta \in \left\lbrace 0.05, 0.1, 0.2, 0.5\right\rbrace $ and make comparisons with the results obtained from Monte Carlo simulations. As seen from Fig. \ref{fig:Fig8}, BER performance of the TherMod scheme is highly dependent on the $\delta$ value, which has an SNR-like effect on the detection mechanism, which is more dominant compared to increasing $N$. We note that the theoretical BEP obtained from \eqref{eq:29} perfectly matches with computer simulation results when the sample variance ($\hat{\sigma}_s^2$) directly generated as a Gaussian random variable, a similar phenomenon reported for the KLJN scheme. However, slight deviations are observed again due to the insufficient fit of chi-square distribution to the Gaussian distribution, particularly with increasing $\delta$.

\begin{figure}[!t]
	\centering
	\includegraphics[width=0.85\columnwidth]{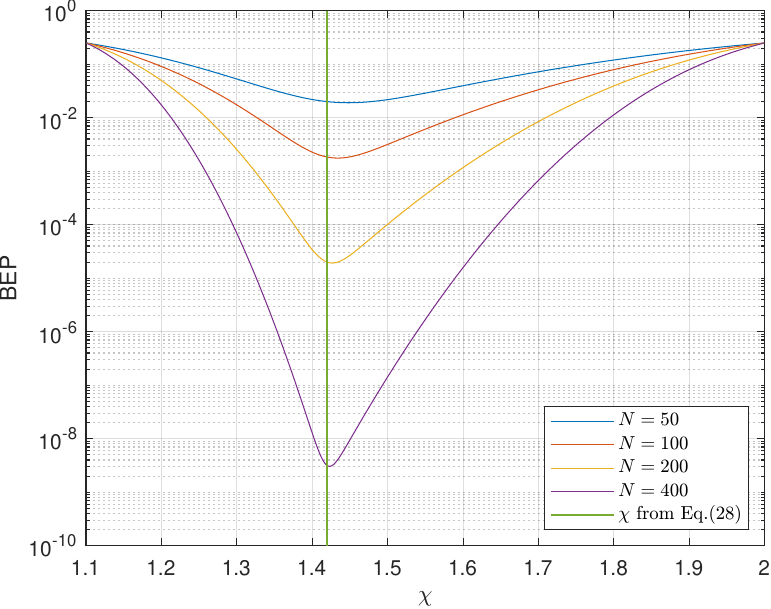}
	\vspace*{-0.3cm}\caption{The search for minimum BEP with respect to the selected threshold value ($\chi$) for varying number of samples ($N$) in TherMod. The vertical line indicates the fixed $\chi$ obtained from \eqref{eq:28}.}\vspace*{-0.3cm}
	\label{fig:Fig9}
\end{figure}

In Fig. \ref{fig:Fig9}, to observe the effect of the selected threshold on the BEP performance, we search for the minimum BEP using \eqref{eq:27} for different $N$ values, where $\delta=0.1$ is assumed. Here, $\chi$ is varied from $1+\delta=1.1$ to $1+\alpha\delta=2$ with increments of $0.001$. For reference, the fixed threshold value obtained from \eqref{eq:28} is also marked in this figure as a vertical line at $\chi=1.4194$. As seen from Fig. \ref{fig:Fig9}, the $\chi$ value that minimizes the BEP slightly changes with respect to $N$. On the other hand, we observe that the fixed $\chi$ value, which ensures a uniform error probability for bit $0$ and bit $1$, provides approximately the minimum BEP for all cases, which can be also verified by substituting numerical $ N $, $\alpha $, and $\delta $ values in \eqref{eq:29}.

\begin{figure}[!t]
	\centering
	\includegraphics[width=0.85\columnwidth]{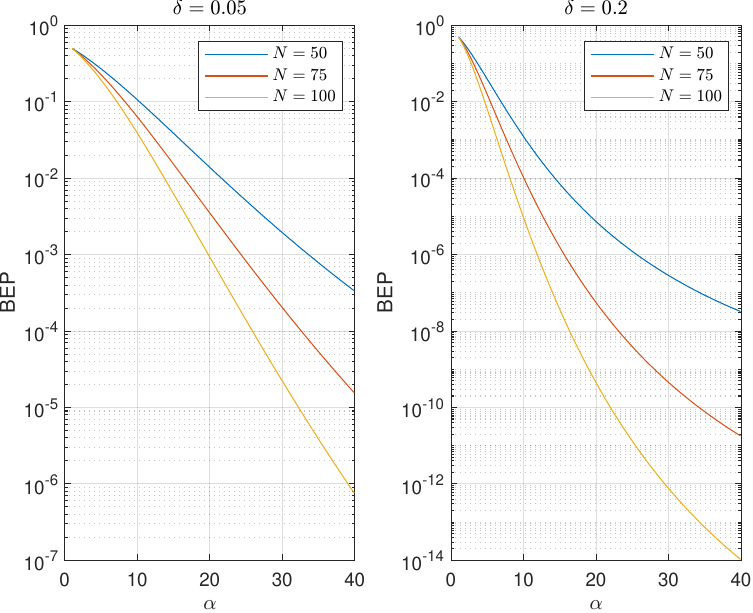}
	\vspace*{-0.3cm}\caption{The effect of ratio ($\alpha$) of resistance loads at the transmitter on BEP of TherMod for two different $\delta$ values.}\vspace*{-0.3cm}
	\label{fig:Fig10}
\end{figure}

In Fig. \ref{fig:Fig10}, we investigate the effect of the ratio ($\alpha$)  of two resistance values at the transmitter on the BEP performance. As discussed earlier, in our basic model, we simply assume that $\alpha$ directly scales the variance of the useful noise samples, as a result, it should be sufficiently high for the reliable separation of two possible noise levels at the receiver. Here, we vary $\alpha$ from $1$ to $40$ for two different $\delta$ values. As seen from Fig. \ref{fig:Fig10}, increasing $\alpha$ directly improves BEP, while the level of saturation is more evident for the case of $\delta=0.2$, where the effect of AWGN samples are less severe. We also note that for $\alpha$ values closer to unity, BEP closely converges to $0.5$. Nevertheless, practical issues related with the selection of higher $\alpha$ values as well as their effect on the receiver side, would be worth of investigation in the future.

\begin{figure}[!t]
	\centering
	\includegraphics[width=0.85\columnwidth]{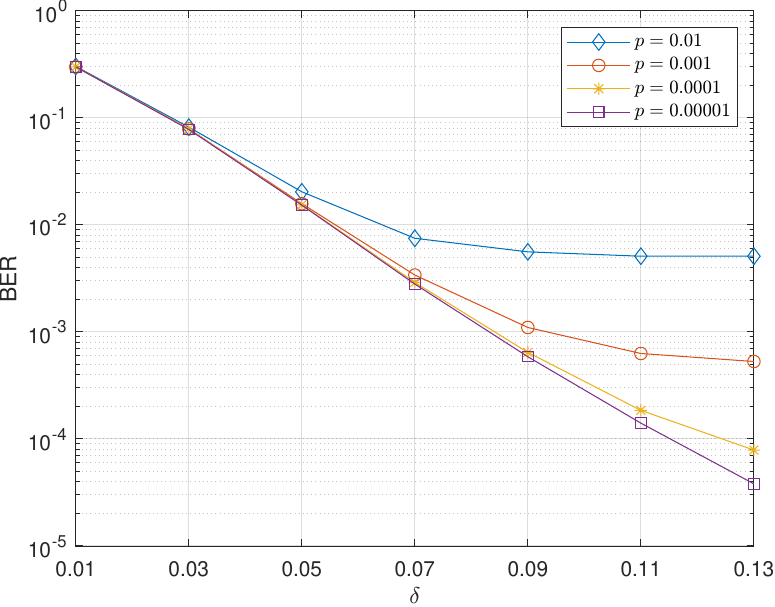}
	\vspace*{-0.3cm}\caption{The effect of useful and disruptive noise variances ($\delta$) on BER in the presence of impulsive noise for TherMod.}\vspace*{-0.3cm}
	\label{fig:Fig11}
\end{figure}

Finally, in Fig. \ref{fig:Fig11}, we investigate the effect $\delta$ on the BER performance by also considering the impulse noise. Here, the impulse noise is modeled as a Bernoulli-Gaussian process in the complex baseband with an impulse probability of $p$ and impulse power that is ten times of the AWGN power \cite{Ghosh_1996}. As seen from Fig. \ref{fig:Fig11}, more frequent impulse noises (higher $p$) cause irreducible error floors by disturbing the threshold-based detection.

\section{Conclusions}
In this paper, we have laid the theoretical fundamentals of communication by means thermal noise, or simply TherCom, for extremely energy-efficient  wired/wireless networks of the future. Particularly, we have put our emphasis on the -not fully perceived- KLJN secure bit exchange scheme, and then taking inspiration from it, we have proposed the wireless TherMod scheme. We note that the power consumption of TherCom schemes can be as low as BackCom systems when operated in the stealth mode by simply indexing the available resistors, while one can expect relatively higher power requirements for the case of external noise generators. We conclude that this preliminary work and its findings might help to unlock the true potential of communication through noise-like signals. The following topics might be of interest for future research: generalization of TherCom schemes for more than two resistors at the communicating parties to transmit higher number of bits, derivation of information theoretical bounds on the data rate and the secrecy capacity, exploration of practical issues such as wire resistances, sampling imperfections, and timing mismatches, and proposal of potential coding schemes to further improve the error performance.

\bibliographystyle{IEEEtran}
\bibliography{IEEEabrv,bib_2022}

\vspace*{-8cm}
\begin{IEEEbiography}[{\includegraphics[width=1in,height=1.25in,clip,keepaspectratio]{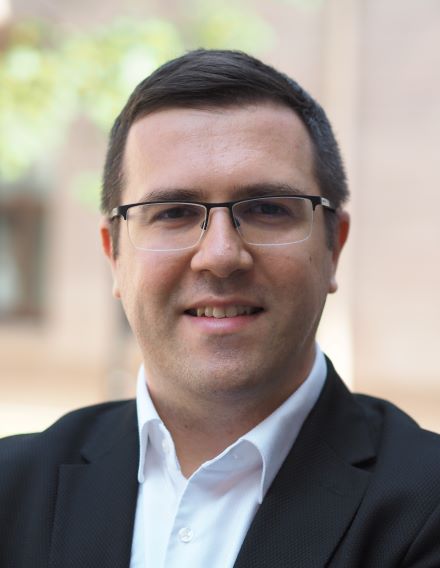}}]{Ertugrul Basar}(S'09-M'13-SM'16-F'23) received his Ph.D. degree from Istanbul Technical University in 2013. He is currently an Associate Professor with the Department of Electrical and Electronics Engineering, Koç University, Istanbul, Turkey and the director of Communications Research and Innovation Laboratory (CoreLab). He had visiting positions at Ruhr University Bochum, Germany (2022, Mercator Fellow) and Princeton University, USA (2011-2012, Visiting Research Collaborator). His primary research interests include beyond 5G and 6G wireless networks, communication theory and systems, reconfigurable intelligent surfaces, index modulation, waveform design, and signal processing for communications. 
	
In the past, Dr. Basar served as an Editor/Senior Editor for many journals including \textsc{IEEE Communications Letters} (2016-2022), \textsc{IEEE Transactions on Communications} (2018-2022), \textit{Physical Communication} (2017-2020), and \textsc{IEEE Access} (2016-2018). Currently, he is an Editor of \textit{Frontiers in Communications and Networks}. He is a Young Member of Turkish Academy of Sciences (2017) and a Fellow of IEEE (2023). 
	
\end{IEEEbiography}

\end{document}